\documentclass[fleqn,usenatbib]{mnras}

\usepackage{newtxtext,newtxmath}
\usepackage{xspace,amsmath}
\usepackage{color,epsfig,amssymb,lscape}
\usepackage{array,colortbl}

\usepackage[T1]{fontenc}
\usepackage{ae,aecompl}

\newcommand{\hii}{\relax \ifmmode {\mbox H\,{\scshape ii}}\else H\,{\scshape ii}\fi}
\newcommand{\mi}{\relax \ifmmode {\mu{\mbox m}}\else $\mu$m\fi}
\newcommand{\ha}{\relax \ifmmode {\mbox H}\alpha\else H$\alpha$\fi}
\newcommand{\hb}{\relax \ifmmode {\mbox H}\beta\else H$\beta$\fi}

\newcommand{\sii}{\relax \ifmmode {\mbox S\,{\scshape ii}}\else S\,{\scshape ii}\fi}
\newcommand{\siii}{\relax \ifmmode {\mbox S\,{\scshape iii}}\else S\,{\scshape iii}\fi}
\newcommand{\nii}{\relax \ifmmode {\mbox N\,{\scshape ii}}\else N\,{\scshape ii}\fi}
\newcommand{\oi}{\relax \ifmmode {\mbox O\,{\scshape i}}\else O\,{\scshape i}\fi}
\newcommand{\oii}{\relax \ifmmode {\mbox O\,{\scshape ii}}\else O\,{\scshape ii}\fi}
\newcommand{\oiii}{\relax \ifmmode {\mbox O\,{\scshape iii}}\else O\,{\scshape iii}\fi}
\newcommand{\neiii}{\relax \ifmmode {\mbox Ne\,{\scshape iii}}\else Ne\,{\scshape iii}\fi}

\newcommand{\rdostres}{\relax \ifmmode {\,\mbox{R}}_{\rm 23}\else \,\mbox{R}$_{\rm 23}$\fi} 

\newcommand{\ciii}{\relax \ifmmode {\mbox O\,{\scshape iii}}\else C\,{\scshape iii}\fi}
\newcommand{\civ}{\relax \ifmmode {\mbox O\,{\scshape iii}}\else C\,{\scshape iv}\fi}
\newcommand{\heii}{\relax \ifmmode {\mbox He\,{\scshape ii}}\else He\,{\scshape ii}\fi}

\newcommand{\gsim}{\hbox{\rlap{\lower.55ex\hbox{$\sim$}} \kern-.3em
\raise.4ex \hbox{$>$}}}
\newcommand{\lsim}{\hbox{\rlap{\lower.55ex\hbox{$\sim$}} \kern-.3em
\raise.4ex \hbox{$<$}}}

\usepackage{graphicx}	
\usepackage{amsmath}	
\usepackage{amssymb}	

\title[Gas-phase C and O abundances in the rest UV]{Using photo-ionisation models to derive carbon and 
oxygen gas-phase abundances in the rest UV}

\author[Enrique P{\'e}rez-Montero \& Ricardo Amor\'\i n]{
Enrique P{\'e}rez-Montero,$^{1}$\thanks{E-mail: epm@iaa.es (EPM)}
Ricardo Amor\'\i n,$^{2,3,4}$
\\
$^{1}$Instituto de Astrof\'\i sica de Andaluc\'\i a. CSIC. Apartado de correos 3004. 18080, Granada, Spain.\\
$^{2}$INAF-Osservatorio Astronomico di Roma, via di Frascati 33, I-00078, Monte Porzio Catone, Italy.\\
$^{3}$Cavendish Laboratory, University of Cambridge, 19 JJ Thomson Avenue, Cambridge, CB3 0HE, UK.\\
$^{4}$Kavli Institute for Cosmology, University of Cambridge, Madingley Road, Cambridge CB3 0HA, UK
}

\date{Accepted XXX. Received YYY; in original form ZZZ}

\pubyear{2016}

\begin{document}
\label{firstpage}
\pagerange{\pageref{firstpage}--\pageref{lastpage}}
\maketitle

\begin{abstract}
We present a new method to derive oxygen and carbon abundances using the ultraviolet (UV) lines emitted by the gas-phase ionised by massive stars. The method is based on the 
comparison  of the nebular emission-line ratios with those predicted by a large grid of photo-ionisation models.   
Given the large dispersion in the O/H - C/O {plane}, our method firstly fixes
C/O using ratios of appropriate emission lines and, in a second step, calculates
O/H and the ionisation parameter from carbon lines in the UV.  
We find abundances totally consistent with those provided by the direct method when 
we apply this method to a sample of objects with an empirical determination 
of the electron temperature using optical emission lines.
The proposed methodology {appears as a powerful tool} for
systematic studies of nebular abundances in star-forming galaxies
at high redshift.
\end{abstract}




\section{Introduction}

 Establishing the gas-phase abundances of carbon, nitrogen and oxygen in galaxies 
through cosmic time is key to understand not only their chemical 
enrichment, but also how galaxies assemble and evolve  
\citep{EdmundsPagel1978, garnett95, henry00, Chiappini2003}. 

{ In the last decade or so, large deep optical spectroscopic surveys using ground-based 
8-10m-class telescopes \citep[e.g.][]{Steidel2003,Shapley2003,zcosmos,Kurk2013,vuds} 
led to a significant increase in the amount and depth of   
rest-frame UV spectra of high redshift galaxies ($z\gtrsim$\,1-2). 
These and other follow-up spectroscopy of galaxies in deep fields \citep[e.g.][]{erb10,Karman2015,Contini2016}, in particular gravitationally lensed systems 
\citep[e.g.][]{christensen12,james14,stark14,Vanzella16}, are pushing now UV 
studies towards the low-luminosity (mass) regime out to 
redshift $z\sim$\,3.5 \citep[e.g.][]{Patricio2016,Vanzella2016b,Vanzella2016c,Amorin16} 
and beyond \citep[][and references therein]{Stark2017}.  }

{ Overall, the above studies consistently find that the 
UV spectrum of high redshift galaxies is systematically bluer and 
harder than in most local galaxies. Thus, it includes a number of
high ionisation, metal emission lines with relative large equivalent
widths, such as those of carbon} (e.g. \ciii] 1908 \AA, \civ 1549 \AA)
  and oxygen (\oiii] 1664 \AA),{ which are often related to the presence of high equivalent width emission of} \heii\,1640 \AA\ and Ly$\alpha$.
    { While these spectral features could be quite common at very high redshifts \citep{Stark2017}, they provide tight constraints to the
      ionisation, age, and metallicity properties of galaxies, as recently 
shown by emission-line diagnostics based on detailed photo-ionisation models  
\citep[e.g.][]{Feltre2016,Gutkin2016,Jaskot2016}. }

{ The wealth of information contained in the UV spectra of strong emission -line galaxies allow us 
to explore new techniques for an accurate characterisation of 
the nebular chemical content of star forming galaxies based on the 
analysis of the carbon abundance, the processes governing the C/O ratio, and 
its relation with the total metallicity \citep[e.g.][]{henry00,erb10,berg16}. 
The use of photo-ionisation models assuming different combinations of
C/O and O/H in an emitting ionised gas distribution is hence an alternative
to estimate the gas-phase metallicity in those cases where very few UV emission lines
can be measured. These methodologies, entirely or partially based on the 
available UV emission line ratios, are necessary to complement the widely 
used recipes to derive physical properties and chemical abundances from 
rest-frame optical spectra, which at $z\sim$\,2-3 rely on the (usually challenging) detection of faint emission lines in the near infra-red \citep[NIR, e.g.][]{Steidel2014,Maseda2014,Troncoso2014,Amorin2014,Shapley2015,Onodera2016,Sanders2016,Trainor2016}. They would also be of great interest for metallicity
studies of galaxies at the highest redshifts, where observing the
optical rest-frame is extremely challenging.
In particular, the future arrival of unprecedentedly deep observations 
using the {\it James Webb Space Telescope} (JWST) and the new 
generation of 30m-class telescopes, will increase by a huge factor current galaxy 
samples at $z\gtrsim$\,3, extending detailed UV spectroscopic studies towards the 
cosmic dawn.}

For this aim, { in this paper} we present an adapted version of the code 
{\sc Hii-Chi-mistry} (\citealt{hcm}, hereinafter {\sc HCm}),
originally designed to derive O/H and N/O from optical emission lines,
that deals with UV emission lines to derive O/H and C/O ratios. { This new 
semi-empirical method is fully consistent with direct estimations based on 
the electron temperature of star forming regions and provide  
a potentially powerful tool to constrain the metallicity of galaxies out to 
very high redshifts. 
Even considering the inherent limitations affecting all methods for 
estimating chemical abundances in galaxies from their integrated spectra, 
such as, e.g. the fact that these are based on spatially unresolved data 
(e.g. \citealt{jip16}) and rely on the simplistic assumption of a single} 
\hii\ { region for the analysis of the observed emission lines, it is still 
possible to find characteristic abundance ratios (i.e. O/H, N/O, \citealt{pm16})
that correlate with the integrated properties of galaxies, such as stellar mass,
thus providing useful constraints for their assembly histories and chemical 
evolution.}

Our paper is organized as follows; In Section~2 we describe the sample
of objects with available emission lines both in the optical and in the UV
whose O and C abundances could be derived using the direct method.
In Section~3 we describe the models and the method used to derive the abundances
and we compare our results with the direct abundances.
Finally, in Section~4 we present our results and in Section 5 we 
summaries our conclusions.

Throughout this paper the following convention is adopted:
O/H is 12+log(O/H), C/O is log(C/O), \oiii]\,$\lambda$\,1664\AA\ represents
the total flux of 1661 and 1666\AA, \ciii]\,$\lambda$\,1908\AA\ represents
the total flux of lines at 1907 and 1909\AA, and \civ\,$\lambda$\,1549\AA\
represents the total flux of lines at 1548 and 1551\AA.

\section{Calibration sample and abundances}

We compiled from the literature a sample of \hii\ regions and starburst 
galaxies at different $z$ with optical and UV lines emitted by the gas 
ionised by episodes of massive star formation. { Owing to the scarce number 
of objects with simultaneous measurement of optical and UV lines, our sample 
consists on a combination of integrated spectra observed using a variety of 
instruments and techniques.}
Our sample was selected in such a way that the O/H and C/O abundance ratios could be derived using the direct method from the estimation of the [\oiii] electron temperature, 
$t_{\rm e}$([\oiii]). This was obtained from the ratio between [\oiii]\,5007 and 
4363\AA\ and, in some cases, from the ratio between [\oiii]\,5007 and 1664\AA.
{ Our calibration sample includes objects without observed fluxes in the} 
[\oii] 3727 \AA\ { line. For these objects the the total oxygen abundance was derived 
by applying an empirical calibration between O/H and} $t_{\rm e}$([\oiii]) {(see below)}.
{ For C/O determinations using the direct method, we compiled objects with 
measurements of at least} \oiii] 1664 \AA\ and \ciii] 1909 \AA, { and also} \civ 1549 \AA\ { in very high-excitation objects.}
All the compiled fluxes were reddening corrected using a \cite{ccm89} extinction law based 
on the extinction coefficients given in these references. 
The list of references and the number of objects 
are listed in Table \ref{table1}.

\begin{table}
\caption{References and number of \hii\ regions, local starbursts (LSB), and Lyman-break galaxies (LBG) used in this work.
}
\begin{center}
\begin{tabular}{lcc}
\hline
\hline
Reference & Number of objects & Type \\
\hline
\cite{deBarros16} & 1& LBG \\
\cite{bayliss14}  &  1  &  LBG \\
\cite{berg16}     &   7  &  LSB \\
\cite{christensen12} & 3  & LBG \\
\cite{erb10}  & 1  &  LBG \\
\cite{garnett95} &  6  &  LSB \\
\cite{garnett97} &  2  &  LSB \\
\cite{garnett99} &  6 &  {\hii} \\
\cite{izotov99} &  3 &  LSB \\
\cite{james14} &  1  & LBG \\
\cite{kurt95} & 2 & {\hii} \\
\cite{kob97}  &  3 & LSB \\
\cite{ks98}  &  3  & LSB \\
\cite{Steidel16} & 1 (stack) & LBG \\ 
\cite{thuan99} & 1  & LSB \\
\cite{Vanzella16} & 1 & LBG \\
\cite{villar04} & 1 & LBG \\
\hline

\end{tabular}
\end{center}
\label{table1}
\end{table}

We recalculated the physical properties and the chemical abundances
of the selected objects using expressions based on the {\sc pyneb} 
software \citep{pyneb}. The O/H abundances were calculated using the 
derived electron temperatures and the bright optical emission lines
 [\oii] {3727 \AA\ and} [\oiii] { 4959,5007 \AA\ and the expressions given
in \cite{hcm}.} 
{ In those cases where the flux of the} [\oii] 3727 \AA\ line were not available 
to calculate the O$^+$ abundances, we resorted to the empirical relation between O/H and $t_{\rm e}$([\oiii]) given by \cite{amorin15}{, which is valid for a wide range of 
metallicity and ionisation values}.

Using the {\sc pyneb} code, we also defined expressions for the temperature and carbon
abundances using UV emission lines. 
Computing the emission line ratio:

\begin{equation}
R_{O3} = \frac{\textrm{I(5007 \AA)}}{\textrm{I(1664 \AA)}}
\end{equation}
we derive the [\oiii] electron temperature using the same set of atomic parameters 
as in \cite{hcm}:

\begin{equation}
\textrm{$t_{\rm e}$([\oiii])} = 1.9263 + 6.433\cdot 10^{-5}\cdot R_{O3} - 0.4221 \cdot \log(R_{O3})
\end{equation}

\noindent valid in the range $t_{\rm e}$([\oiii]) = 0.8 to 2.5 in units of 10$^4$ K 
and that is nearly independent on electron density
{ (i.e. a change from 10 to 1000 cm$^{-3}$ implies a change in $t_{\rm e}$ of
less than 0.1\%).}
{ Nevertheless, due to the very long wavelength baseline between the involved emission
lines and the large extinction correction in the UV, the used ratio has a large uncertainty
if the extinction correction is big. For instance assuming an extinction of 1 mag
at 1664 \AA\  implies an electron temperature of about 10\% larger using a \cite{calzetti00} than a \cite{ccm89} extinction law.}

The carbon ionic ratios based on UV lines were derived using the 
following expressions and assuming the derived $t_{\rm e}$[\oiii]:

\[\log(C^{2+}/O^{2+}) = \log\left(\frac{\textrm{\ciii] 1908 \AA}}{\textrm{\oiii] 1664 \AA}}\right) - 0.8361 + \]
\begin{equation}
- \frac{0.4801}{t_e} - 0.0358 \cdot t_e + 0.2535 \cdot \log(t_e)
\end{equation}

\[\log(C^{3+}/O^{2+}) = \log\left(\frac{\textrm{\civ 1549 \AA}}{\textrm{\oiii] 1664 \AA}}\right) - 1.5625 + \]
\begin{equation}
+ \frac{0.2946}{t_e} - 0.043 \cdot t_e + 0.31 \cdot \log(t_e)
\end{equation}
and using the collisional coefficients of \cite{c2+} and \cite{c3+} for
C$^{2+}$ and C$^{3+}$, respectively.

The total C/O was then derived under the assumption that:

\begin{equation}
\frac{C}{O} = \frac{C^{2+}+C^{3+}}{O^{2+}}
\end{equation} 

\noindent that can be simplified to C/O = C$^{2+}$/O$^{2+}$
when no \civ\ emission is detected.

\begin{figure}
\centering
\includegraphics[width=8cm]{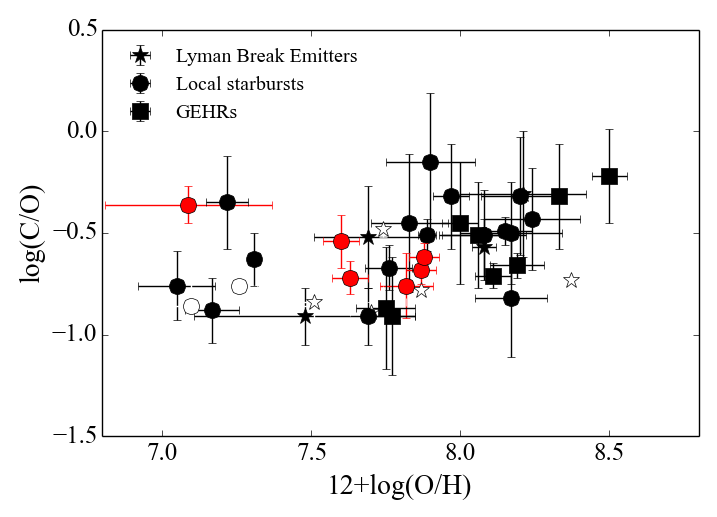}
\caption{Relation between O/H and C/O for the studied sample
as derived from the direct method. { Different
symbols stand for the class of objects and colours for the method
used to derive abundances: direct method with 5007/4363 ratio (black), 
O/H from empirical relation with} $t_{\rm e}$[\oiii] { (red), and 
direct method using temperature from 5007/1664 ratio (white).}
}
\label{oh-co}
\end{figure}

In Fig.\ref{oh-co} we plot the relation between total O/H and C/O for
our calibration sample. 
{ The objects are identified according to their nature and the method used
to derive O/H and C/O. We distinguish between objects with temperatures obtained 
from the} [\oiii] { ratio 5007/4363 (30), objects whose O/H was calculated
using the empirical relation between O/H and} $t_{\rm e}$([\oiii]) { given in \cite{amorin15} (6), and objects whose O/H and C/O were calculated using the 
electron temperature from the} [\oiii] { 5007/1664 ratio (8). 
For the latter, typical errors in O/H are $\sim$\,0.18 dex, while in the two 
first cases they are slightly lower, $\sim$\,0.10 dex.}

On average, C/O tend to increase with metallicity, as carbon 
is essentially a secondary element (i.e. its relative production is higher 
for higher O/H) mainly ejected into the ISM by massive stars 
(e.g. \citealt{henry00}). 
{ This trend starts to be observed in our sample from 12+log(O/H)\,$>$\,7.4.}
However, the dispersion of this relation, { of about 0.25 dex,  is larger than  
 the average C/O errors ($\sim$\,0.15 dex),} 
owing to the different physical processes affecting metallicity and the ratio
between a primary element as O and a secondary one as C. 
This is the case of hydrodynamical processes, including metal-enriched 
outflows or inflows of metal-poor gas, which can change 
metallicity and keep relatively unaffected C/O
(e.g. \citealt{edmunds90}). Other factors that could also 
modify this ratio are, for instance, different star-formation efficiencies 
or the presence of a non-standard IMF (e.g. \citealt{gavilan05,Mattsson10})
Therefore, as no clear { trend} can be taken between O/H and C/O, 
C/O has to be previously calculated to allow the derivation of 
O/H abundances  using carbon lines.
This is similar to the situation in the optical, when N/O must be
determined before deriving O/H using [\nii] emission lines
\citep{pmc09,amorin10}.

\section{Model and method description}

Both O/H, C/O and the ionisation parameter (log$U$) were derived 
using a semi-empirical approach based on the comparison between the 
most representative extinction corrected emission lines observed in 
the UV (Ly$\alpha$\,1216\AA, \civ\,1549\AA, \oiii]\,1664\AA, and 
\ciii]\,1908\AA) and in the optical (\hb\,4861\AA\ and [\oiii]\,
5007\AA) with  a large grid of photo-ionisation models covering 
a wide range of values in O/H, C/O, and log$U$.

The photo-ionisation models were calculated using the code 
{\sc Cloudy} v.13.01 \citep{cloudy}. In all models we adopted 
a spherical geometry of constant density gas (100 cm$^{-3}$), 
filling factor of 10$^{-1}$, and a standard gas-to-dust mass 
ratio around an ionising point-source with the spectral energy 
distribution of a 1 Myr old cluster using {\sc popstar} 
\citep{popstar}  model atmospheres at the metallicity of the 
gas. The calculation was stopped in each model and the lines 
were thus retrieved when 
the electron density fraction was lower than 10\%.

{ Although several authors have pointed out the importance of
the assumed SED in the models for the determination of abundances
(e.g. \citealt{kewley05, morisset16}), as our approach is
semi-empirical it is possible to assess any systematic effect and
the uncertainty associated to the assumption of an ionizing
stellar cluster of a single age in our models.}

The grid { covers} input oxygen abundances in the range [7.1,9.1] in 
bins of 0.1 dex, log$U$ in the range [-4.0,-1.5] in bins of 0.25 dex and
assumes solar proportions for the rest of elements, with 
the exception of carbon, that takes values in the range 
of C/O = [-1.4,0.6] in bins of 0.125 dex, and nitrogen, 
for which we assume a secondary origin and a constant solar 
ratio log(C/N)$=$\,0.6  \citep{asplund}. 
{ The assumption of a constant C/N ratio is justified by the  
expected similar behavior between primary and secondary elements 
under hydrodynamical effects (e.g. \citealt{edmunds90}) and 
supported by recent observations of galaxies at low  
\citep{berg16} and high redshifts \citep{Steidel16}, which show a constant  
trend in C/N with metallicity. Under the above conditions, the grid consists on} 
a total of 3,927 models that can be obtained from the 3MDB database \citep{3mdb}
\footnote{3MdB project and the models can be found in \tt{https://sites.google.com/site/mexicanmillionmodels/}}.

A Python-based routine called {\sc Hii-Chi-mistry-UV} \footnote{The routine
{\sc HCm-UV} is publicly available at {\tt http://www.iaa.es/$\sim$epm/HII-CHI-mistry-UV.html}.}
(hereafter {\sc HCm-UV} was developed in order to
calculate O/H, C/O and log$U$ comparing the 
observed relative UV and optical emission lines with the predictions from 
the models. The methodology is the same as that described for the optical 
in \cite{hcm}. In a first step, the code calculates C/O in order to constrain the space
of models in which O/H abundances can be calculated using UV carbon emission lines.

\begin{figure*}
\centering
\includegraphics[width=8cm,clip=]{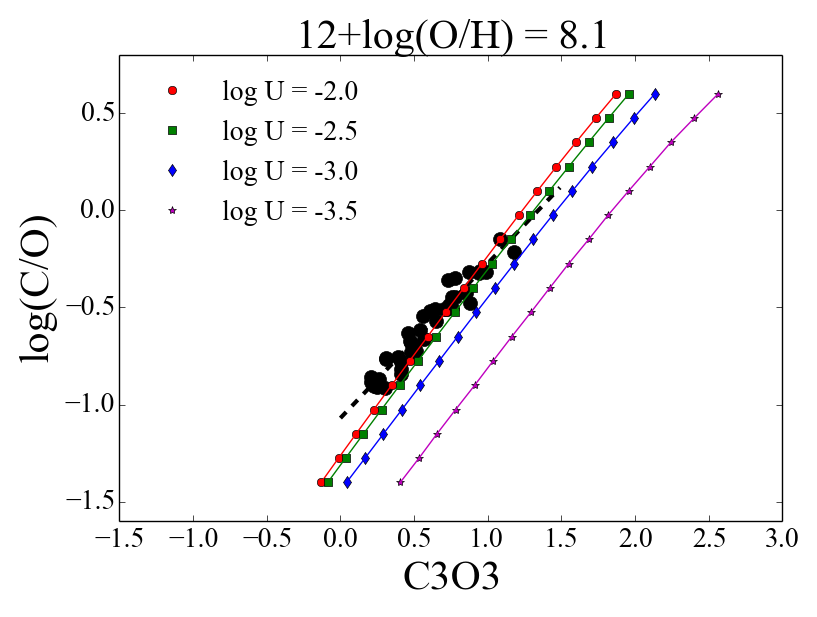}
\includegraphics[width=8cm,clip=]{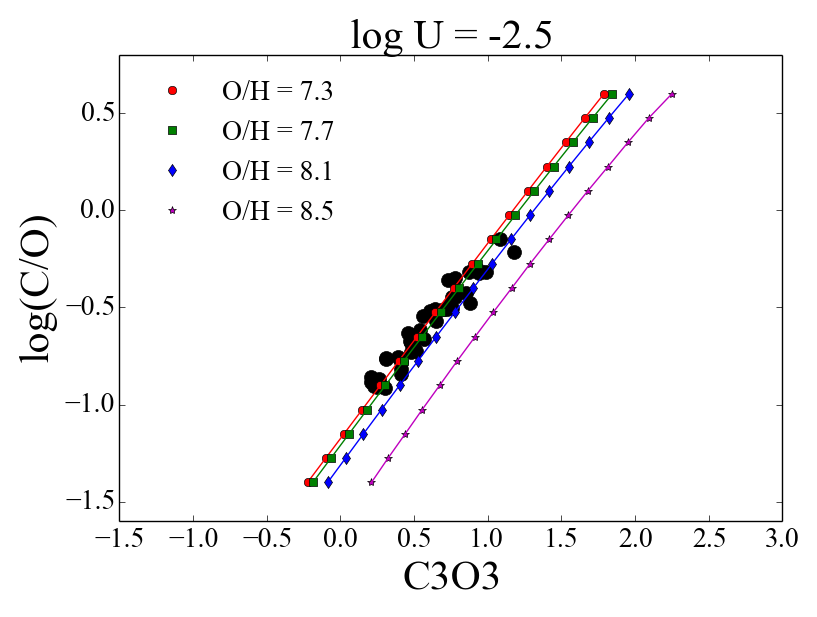}
\includegraphics[width=8cm,clip=]{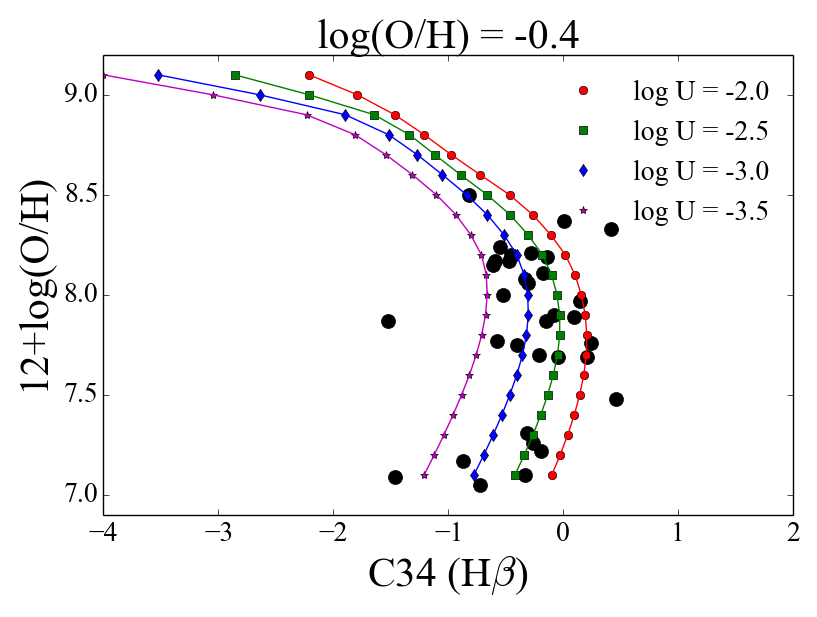}
\includegraphics[width=8cm,clip=]{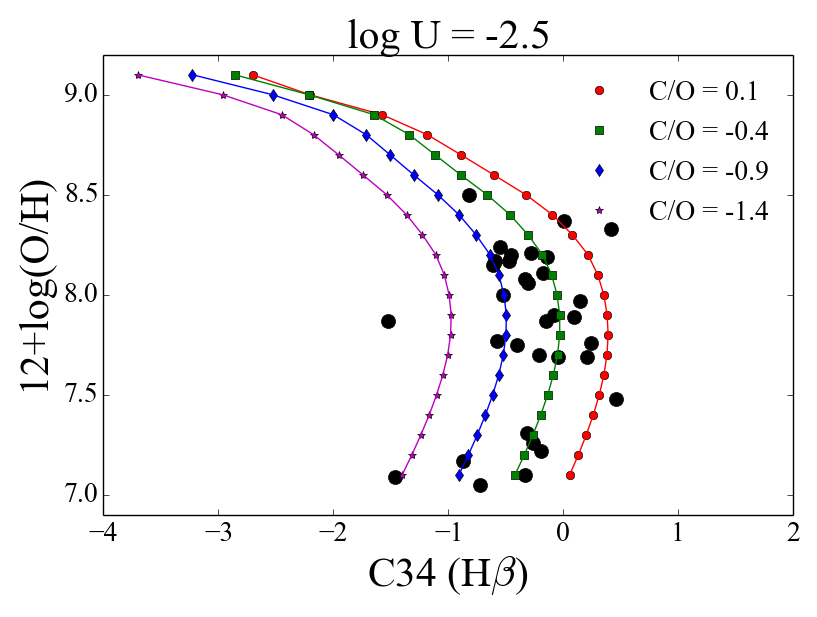}
\caption{Relations between the relevant emission line ratios, models, and derived 
abundances for the compiled sample. Upper panels: Relation between C3O3 and C/O at fixed O/H (Left)
and fixed log$U$ (Right).{ The dashed black line represents the linear
fitting to the observations.}
Lower panels: Relation between C34 and O/H at fixed C/O
(Left) and at fixed log$U$ (Right). }
\label{mods}
\end{figure*}

The C/O abundance ratio is derived as the average of the $\chi^2$-weighted 
distribution of the C/O values in the models. { The $\chi^2$ values are 
calculated as the relative quadratic differences between observed and
model-predicted emission-line ratios sensitive to C/O.
The ratio R$_{O3}$ is used when} \oiii] { 1664 and
5007 \AA\ are observed. In all cases, we also use the 
$C3O3$ parameter, defined as:}

\begin{equation}
C3O3 = \log\left(\frac{\textrm{I(\ciii] 1908 \AA) + I(\civ 1549 \AA)}}{\textrm{I(\oiii] 1664 \AA)}}\right)
\end{equation}

\noindent which can be also defined in absence of the \civ\ line 
as the ratio of the \ciii] and [\oiii] lines. 
As shown in Fig.\ref{mods}, according to models $C3O3$ has little dependence 
on O/H, { while the dependence on log$U$ is only significant for 
very low-excitation. 
Figure~\ref{mods} also shows the calibration sample, which follow a
linear trend that can be used to derive C/O empirically:}

\begin{equation}
log(C/O) = -1.069 + 0.796 \cdot C3O3
\end{equation}
\noindent { with a dispersion of 0.20 dex, calculated as
the standard deviation of the residuals.}

Once C/O is fixed in the grid of models, both O/H and log$U$ are calculated as
the mean of the model input values in the $\chi^2$-weighted distribution, where again
the $\chi^2$ values for each model are derived from the quadratic relative differences
between the observed and predicted { emission-line ratios sensitive to O/H and log U. 
Also in this case, R$_{O3}$ is used when} \oiii]\,1664 and [\oiii]\,5007 { are observed. 
Similarly to the case of C/O, the $\chi^2$ values for $R_{O3}$ are 
added in quadrature to other emission-line ratios from the UV, such as the $C34$
parameter, defined as:}

\begin{equation}
C34 = \log\left(\frac{\textrm{I(\ciii] 1908 \AA) + I(\civ 1549 \AA)}}{\textrm{I(H{\sc i})}}\right)
\end{equation}

\noindent { where I(H{\sc i}) is the intensity of a Hydrogen recombination line. 
 When optical emission lines are available, the HI line is 
 H$\beta$ 4861\,\AA. As in the case of $C3O3$, $C34$ }can be defined in the absence
of \civ\ lines using only \ciii] lines.
As shown in Fig.\ref{mods}, this ratio has a very similar dependence with
O/H as the R$_{23}$ parameter (i.e. [\oii]+[\oiii] relative to \hb, \citealt{r23}) as it
is double-valued (i.e. it increases with O/H for low O/H and it decreases 
for high O/H). Besides, it presents additional dependence on log$U$
and C/O that contribute to enlarge the dispersion. 
The dependence on C/O is limited in our method because 
C/O is constrained in the first iteration of our calculations.
{ As shown in Fig.~\ref{mods}, the $C34$ parameter also presents a 
large dependence on log U. The latter is obtained in this iteration} as the 
average value of the resulting distribution. 
The dependence on log$U$ { is alleviated} using { an additional parameter} in 
the total $\chi^2$ that depends on { the $C3C4$ index, defined as the} ratio between 
the \ciii] and \civ\ lines:

\begin{equation}
C3C4 = \log\left(\frac{\textrm{I(\ciii] 1908 \AA)}}{\textrm{I(\civ 1549 \AA}}\right)
\end{equation}

\begin{figure*}
\centering
\includegraphics[width=8cm]{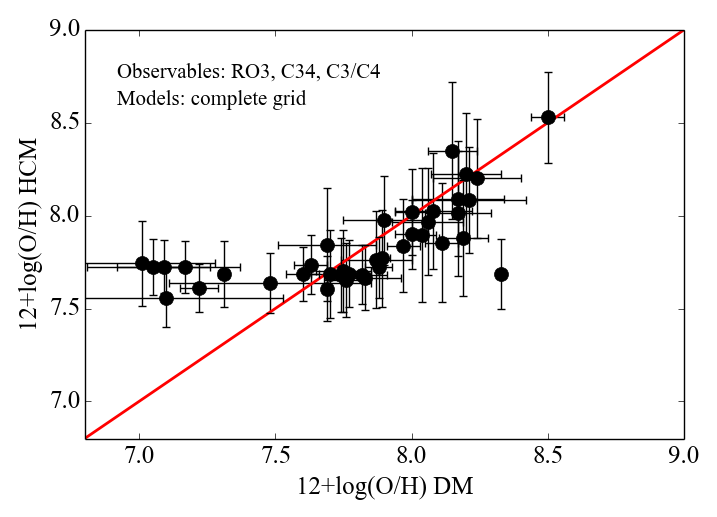}
\includegraphics[width=8cm]{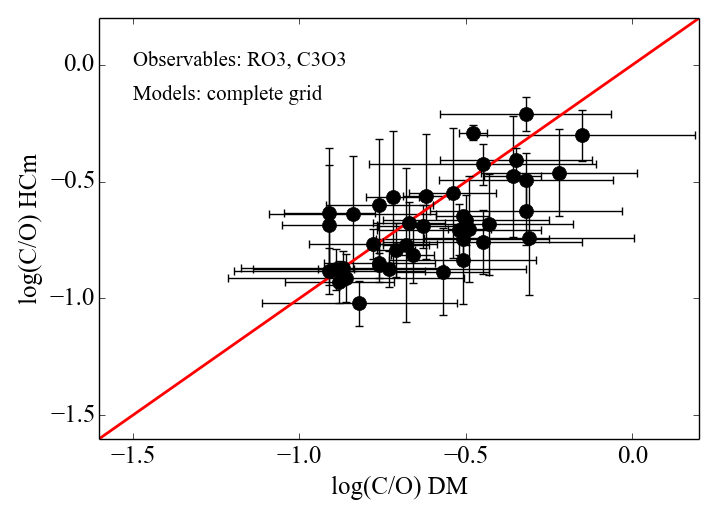}
\includegraphics[width=8cm]{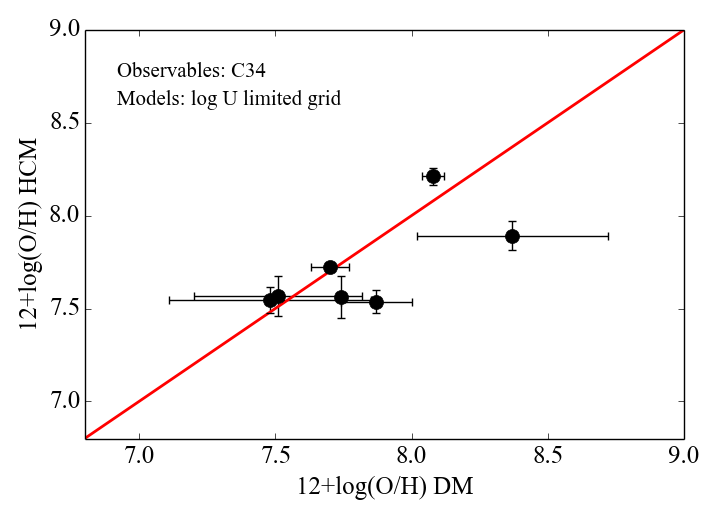}
\includegraphics[width=8cm]{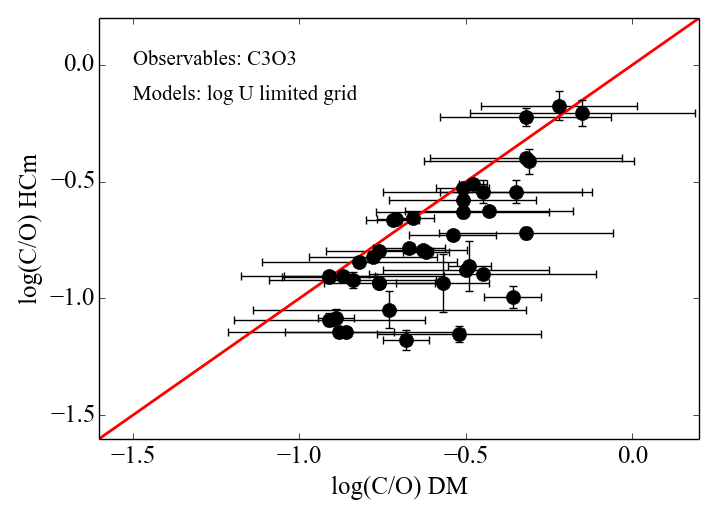}
\caption{Comparison between C/O (Right) and O/H (Left) abundances 
derived from the direct method and from {\sc HCm-UV} under
different assumptions. The {\sc HCm-UV}-based abundances in lower panels were derived using
only UV lines, while for upper panels they also include optical
information, i.e. H$\beta$ and [\oiii] 5007 \AA. 
The red solid lines represent in all cases the 1:1 relation.}
\label{comp}
\end{figure*}

\begin{figure}
\centering
\includegraphics[width=8cm]{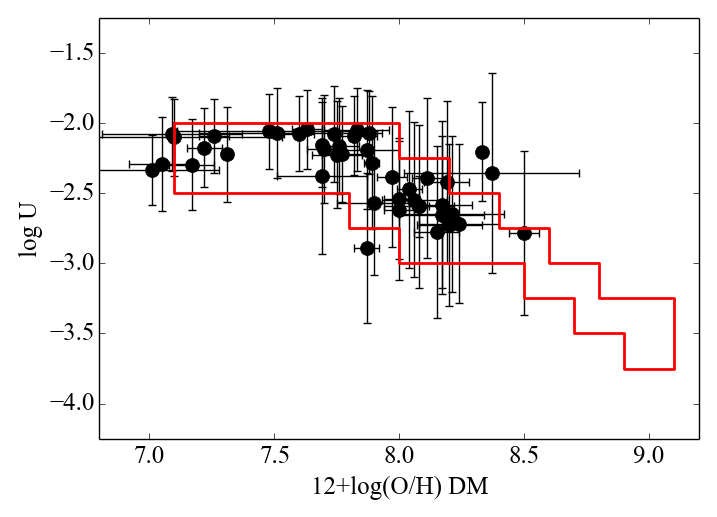}
\caption{{ Relation between O/H as derived from the direct method and
log U derived using {\sc HCm-UV} using all available lines and the complete grid
of models for the selected sample of objects. The red soli line encompasses the grid
of models empirically constrained in P\'erez-Montero (2014).}}

\label{oh-u}
\end{figure}

 \section{Results and discussion}

In Figure~\ref{comp} we show the comparison 
between the chemical abundances derived from both the
direct method and from {\sc HCm-UV}. In the two upper panels,  
the latter uses { all} the available optical (\hb\ and [\oiii]\,5007) 
and UV (\civ, \ciii], \oiii]) emission lines. 
Overall, we find a good agreement between these two datasets,
{ both for C/O and O/H\,$>$\,7.5,} 
with an mean offset { that is} lower than the uncertainty of the 
direct method (i.e. $<$\,0.10 dex) both for O/H and C/O.
The overall standard deviation of the residuals is 0.15 dex for O/H 
and 0.16 dex for C/O.

Interestingly, using {\sc HCm-UV} it could be possible to obtain abundances from
UV lines adding only follow-up observations of the { blue part of the
rest-frame optical} spectrum corresponding to [\oiii] 5007 \AA\ and \hb, without
any low excitation emission line such as {\oii] 3727 \AA.
{ Though for very low values of O/H ($\lesssim$\,7.5) the abundances
derived by {\sc HCm-UV} do not follow any correlation with those
from the direct method, the code essentially identifies all of these
objects as extremely metal-poor galaxies.}

When the $R_{O3}$ ratio --used to derive $t_{\rm e}$([\oiii]) 
in the direct method-- cannot be derived due to the absence of 
optical lines, the code follows the same strategy as in \cite{hcm}: 
it constrains the grid of models by assuming an empirical law 
between O/H and log$U$ in such a way that metal-poor objects have 
higher log$U$ and metal-rich objects have lower log$U$. This 
empirical relation was already observed by \cite{de86} and could 
be related with an evolutionary sequence. Since our sample small 
compared with the sample studied in \cite{hcm} we adopt the same 
constrain for our grid. { In Fig. \ref{oh-u} we show the 
comparison between the empirical O/H - log U relation 
empirically derived for \cite{almeida16} and \cite{pm16} and the values derived for our
sample using all lines. As can be seen the trend is well followed
by the points, }

By using this constrained grid, the code calculates C/O using only the $C3O3$ 
ratio and the same procedure as described above and then derives O/H 
and log$U$ using the $C3C4$ and $C34$ parameters, but 
taking Ly$\alpha$ as the hydrogen recombination line. 
This approach, however, should be used with caution. The Ly$\alpha$ line is 
resonant and photons scatter in the neutral Hydrogen, thus having a 
complex radiation transfer. Especially for faint  Ly$\alpha$ emission 
(i.e. low equivalent width) this may produce an additional source of 
systematic uncertainty.  

In the two lower panels of Fig. \ref{comp} we show the comparison between the 
abundances derived from the direct method and from {\sc HCm-UV} in absence 
of the optical emission lines. The number of points with a reliable measurement 
of Ly$\alpha$ in our sample is very low, but all the objects are identified as 
metal-poor in this sample. Regarding C/O, as this ratio has a very 
little dependence on the electron temperature, the agreement is still very 
good, { i.e. the mean offset is lower than the
uncertainty of the direct method and the standard deviation of the
residuals is 0.20 dex, fairly similar to that obtained 
using optical lines}.

This result underlines the importance of the C/O abundance ratio as an
indicator of the chemical content of galaxies at different epochs. 
Apart from the reliability of the results only using \oiii] and \ciii]
UV emission lines, C/O is not affected by hydrodynamical
processes, such as inflows or outflows, and can be more tightly
related with other integrated properties of galaxies such as stellar
mass, as in the case of N/O \citep{amorin10,pm13, pm16}.

When no previous determination of C/O can be made 
(i.e. the C3O3 ratio cannot be measured), we follow again the procedure described in
\cite{hcm} for the case of N/O: the code adopts a new constrain for the space of
models assuming a given empirical relation between O/H and C/O to calculate O/H using
the C34 parameter. The code then uses the relation between O/H and N/O shown in Fig.~3 of \cite{hcm} and assumes the C/N solar ratio for all the sample.

\section{Conclusion}

{ In this work we present a semi-empirical method 
to derive oxygen (O/H) and carbon abundances (C/O) from  
rest-frame UV emission lines, which are consistent with those obtained using 
the direct method. }

{ We highlight the potential use of this method} for ($T_e$-consistent) 
metallicity inferences at high redshifts ($z>2$), { e.g. through deep optical 
spectroscopy with large ground-based telescopes \citep{Amorin16} or using future 
space telescopes, such as the {\it James Webb Space Telescope}.  
For example, the NIRSpec spectrograph onboard the JWST, will be able
to detect the full set of rest UV and optical lines required to use
this method for galaxies spanning from $z\sim$\,3.5 to $z\sim$\,9
in relatively short exposure times.
Particularly interesting could be the use of this technique to provide 
metallicity constraints for primeval galaxies in the epoch of cosmic
reionisation (at $z>$\,6). 
In these presumably metal-poor objects,} rest UV emission 
lines with relatively high EWs { (such as} \ciii], \civ, and \oiii]) are 
    apparently ubiquitous \citep[e.g.][]{stark15, Stark2017}, but the
    access to faint 
key optical lines (such as [\oiii] 4363 \AA\ and [\oii] 3727 \AA\ 
in the case of high ionisation objects) { will be very limited, thus making} 
$T_e$-consistent metallicity inferences extremely challenging. 

\section*{Acknowledgements}
{ We thank the anonymous referee for constructing and helpful comments.}   
EPM acknowledges support from the Spanish MICINN through grants
AYA2010-21887-C04-01 and AYA2013-47742-C4-1-P and the Junta de 
Andaluc\'\i a for grant EXC/2011 FQM-7058. RA acknowledges the support 
from the ERC Advanced Grant 695671 ``QUENCH'' and the FP7 SPACE  
project ``ASTRODEEP'' (Ref.No: 312725), supported by the European 
Commission.












\bsp	
\label{lastpage}
\end{document}